\documentclass[preprint2]{aastex}

\shorttitle{PRD models: Detonation}
\shortauthors{Bravo \& Garc\'\i a-Senz}

\def\etal{et al.}

\begin{document}

\title{Pulsating reverse detonation models of Type Ia supernovae. I: Detonation ignition}
\author{Eduardo Bravo\altaffilmark{1,2}, 
Domingo Garc\'\i a-Senz\altaffilmark{1,2}}
\altaffiltext{1}{Dept. F\'\i sica i Enginyeria Nuclear, Univ. Polit\`ecnica de
Catalunya, Diagonal 647, 08028 Barcelona, Spain;   
eduardo.bravo@upc.edu, domingo.garcia@upc.edu}
\altaffiltext{2}{Institut d'Estudis Espacials de Catalunya, Barcelona, Spain}

\begin{abstract}
Observational evidences point to a common explosion mechanism of Type Ia supernovae based on a delayed detonation of a white dwarf. Although several scenarios have been proposed and explored by means of one, two, and three-dimensional simulations, the key point still is the understanding of the conditions under which a stable detonation can form in a destabilized white dwarf. One of the possibilities that have been invoked is that an inefficient deflagration leads to the pulsation of a Chandrasekhar-mass white dwarf, followed by formation of an accretion shock around a carbon-oxygen rich core. The accretion shock confines the core and transforms kinetic energy from the collapsing halo into thermal energy of the core, until an inward moving detonation is formed. This chain of events has been termed Pulsating Reverse Detonation (PRD). In this work we explore the robustness of the detonation ignition for different PRD models characterized by the amount of mass burned during the deflagration phase, $M_\mathrm{defl}$. The evolution of the white dwarf up to the formation of the accretion shock has been followed with a three-dimensional hydrodynamical code with nuclear reactions turned off. We found that detonation conditions are achieved for a wide range of $M_\mathrm{defl}$. However, if the nuclear energy released during the deflagration phase is close to the white dwarf binding energy ($\sim0.46\times10^{51}~\mathrm{erg} \Rightarrow M_\mathrm{defl}\sim0.30$~M$_\odot$) the accretion shock cannot heat and confine efficiently the core and detonation conditions are not robustly achieved. 
\end{abstract}


\section{Introduction}

Type Ia supernovae (SNIa) are the most energetic transient phenomena in the 
Universe displaying most of its energy in the optical band of the 
electromagnetic spectrum, where they rival in brightness with their host 
galaxies during several weeks. The  importance of SNIa in astrophysics and 
cosmology is highlighted  by their use as standard (or, better, calibrable) 
candles to measure cosmic distances and related cosmological parameters \citep{per98,sch98,rie01}. 
However, in order to achieve a high precision in the distance determination as 
required, for example, to determine the equation of state of the dark energy 
component of our Universe, it is necessary to understand the physics of SNIa 
explosions. From the theoretical point of view, the accepted model of SNIa 
consists of a carbon-oxygen white dwarf (WD) near the Chandrasekhar mass that accretes matter from a companion in a close binary system. This 
model accounts for the SNIa sample homogeneity, the lack of prominent hydrogen lines in 
their spectra, and its detection in elliptical galaxies. Massive WDs
 are extremely unstable bodies in which a modest release of energy can produce 
 a huge expansion (i.e. their ratio of binding to gravitational energy is 
 only $\sim15\%$, while in a normal star it is $\sim50\%$). There are two main 
 ingredients of the standard model that are still poorly known: the precise 
 configuration of the stellar binary system and its evolution prior to thermal 
 runaway of the WD \citep{lan00,no03,pie03,han04,bad07,hkn08}, and the explosion mechanism \citep{hil00}. Fortunately, as long as 
 a carbon-oxygen WD reaches the Chandrashekar mass, its previous evolution is not critical for the explosion because the WD structure is determined by the state of a degenerate 
 gas of fermions that can be described with a single parameter,  
 the mass of the star. This fact leaves the explosion mechanism as the most 
 relevant unknown concerning SNIa.
 
In spite of continued theoretical efforts dedicated to understand the 
mechanism behind SNIa, realistic simulations are still unable to provide a 
satisfactory description of the details of these thermonuclear explosions. Nowadays, there is a
consensus that the initial phases of the explosion involve a subsonic 
thermonuclear flame (deflagration), whose propagation competes with the 
expansion of the WD.  After a while the corrugation of the flame front  
induced by hydrodynamic instabilities culminates in an acceleration of the 
effective combustion rate. However, the huge difference of lengthscales between the white dwarf ($\sim10^8$~cm) and the flame width ($\lesssim1$~cm) prevents that large-scale numerical simulations resolve the deflagration front, making it necessary to implement a subgrid model of the subsonic flame. The precise value of the effective combustion rate is 
currently under debate, as it depends on details of the flame model.  
Recent three-dimensional calculations by different groups have shown that 
pure deflagration models always give final kinetic energies that fall 
short of $10^{51}$ ergs, while leaving too much unburnt carbon and oxygen close
to the center \citep{rei02b,gam03,gar05,rop07c}. Because both signatures are at odds with observational constraints,  \citet{gam03,gam04,gam05} concluded that the only way to reconcile three-dimensional simulations with observations is to assume that a detonation ignites after a few tenths of a solar mass have been incinerated subsonically (delayed detonation). 

The idea that a delayed detonation is somehow involved in thermonuclear explosions leading to SNIa has been around for many years. 
\citet{iva74} proposed that a delayed detonation might take place after a long-range pulsation of the white dwarf, in what is known as the Pulsating Delayed Detonation (PDD) scenario. \citet{kho91} carefully computed one-dimensional delayed detonation models that were successfully compared with observations of SNIa explosions \citep{hoe95,how01,qui07,ger07,fes07} and their remnants \citep{bad06,bad08b,res08}. In these one-dimensional models the detonation initiation had to be postulated, as there was not identified any sound physical mechanism by which such a transition could happen in an unconfined white dwarf \citep{nie99}. 

Ignition of a detonation in a white dwarf can happen in two basic ways, either through a deflagration-to-detonation transition (DDT) or because of a sudden energy release in a confined fluid volume (hereafter: confined detonation ignition or CDI).
The essential feature of detonation initiation is the
formation of a non-uniformly preheated region with a level of fluctuations of
temperature, density, and chemical composition such that a sufficiently large
mass burns before a sonic wave can cross it \citep{kho91,nie97}. The thermal
gradient needed is \citep{bk87,woo90,kho91}:

\begin{equation}\label{eq1}
\nabla T < \frac{\Theta T}{A v_\mathrm{sound} \tau_\mathrm{i}}
\end{equation} 

\noindent where $A$ is a numerical coefficient, $\tau_\mathrm{i}=T/\dot{T}$ is the
induction timescale at temperature $T$, and $\Theta\sim0.04-0.05$ is the
Frank-Kameneetskii factor:

\begin{equation}
\Theta = - \frac{\partial\ln T}{\partial\ln\tau_\mathrm{i}}
\end{equation}

Such fluctuations could be produced by a variety of mechanisms: adiabatic
pre-compression in front of a deflagration wave, shock heating, mixing of hot
ashes with fresh fuel \citep{kho91}, accumulation of pressure waves due to a
topologically complex geometrical structure of the flame front \citep{woo94}, or transition to
the distributed burning regime \citep{nie97}. Among the proposed 
mechanisms of DDT, turbulence pre-conditioning has received the most attention. \citet{kho97} determined criteria for a DDT in unconfined conditions, such
as those realized during the expansion of a white dwarf following a deflagration
phase. For a DDT to be feasible, the turbulent velocity has to exceed the
laminar flame velocity by a factor $\lesssim8$ at a lengthscale comparable to the
detonation wave thickness. This criteria was fulfilled for flame densities in
the range $5\times10^6$~g/cm$^3 <\rho < 2-5\times10^7$~g/cm$^3$ for reasonable
assumptions. At densities in excess of $10^8$~g/cm$^3$ a DDT is quite
unlikely \citep[but see \citeauthor{zin07} who pointed to bubbles fragmentation as a 
way to increase the flame surface and facilitate a DDT at
$\rho\sim2\times10^8$~g/cm$^3$]{kho97}. 

Small-scale simulations are needed to ascertain if the necessary conditions for
a DDT are actually achieved during a white dwarf explosion driven by a
deflagration wave. Up to the present, such studies seem to disfavor a DDT in
view of the robustness of subsonic flames against interaction with vortical
flow \citep{rop04} for the maximum expected velocities at the
integral scale, $\sim100$~km/s \citep{lis00}. Moreover, \citet{nie99} estimated that the size of the fluctuations that can be expected from turbulence at the critical densities is $\sim3$ orders of magnitude smaller than
required for a DDT to happen. 

Three-dimensional numerical simulations of white dwarfs undergoing a slow deflagration have provided new ways to achieve a CDI. \citet{pl04} followed the evolution of a white dwarf assuming ignition in a single slightly off-centered bubble, and found that a CDI might result from the convergent flow at the antipodes of the initial ignition point. In their model, the detonation was possible because of the confining gravitational field, hence it was termed Gravitationally Confined Detonation \citep[GCD, see also][]{rwh07,pl07,tow07,jor07}. \citet{bra05b} described another CDI scenario in which a detonation was triggered by inertial confinement due to an accretion shock born around a carbon-oxygen white dwarf core after pulsation caused by burning of $\sim 0.1-0.3$~M$_\odot$: the Pulsating Reverse Detonation (PRD)\footnote{A conceptually similar model was proposed by \citet{dun01}. See also \citet{dun03}}. In this work, we explore the robustness of detonation ignition conditions in the PRD scenario, as a function of the mass burned during the deflagration phase prior to pulsation, $M_\mathrm{defl}$. The evolution of the white dwarf up to the formation of the accretion shock has been followed with a three-dimensional hydrodynamic code with nuclear reactions turned off in order to determine the most extreme conditions achieved by the white dwarf during the re-collapsing phase of the pulsation. In the next section we summarize the main features of the proposed scenario, and compare PRD models with one-dimensional PDD models. After that, we analyze if detonation conditions are achieved in our PRD models. We then discuss the implications that the presence of a small cap of helium on the white dwarf at the moment of thermal runaway might have for CDI. Finally, we present the conclusions of this work. A companion paper (Bravo et. al., 2009; hereafter paper~II) is devoted to present the detailed evolution of PRD models after  detonation ignition and their final outcomes.

\section{Briefing of the pulsating reverse detonation model}

In the PRD scenario the explosion proceeds in three steps \citep{bra06}: 1) an initial 
pre-conditioning phase of the WD whose result is an expanded structure with a positive gradient of the mean chemical weight, 2) formation of an accretion shock that confines the
fuel volume and, 3) launch of a reverse, converging, detonation wave. We have
probed the PRD concept by performing three-dimensional simulations of the explosion. The initial model consists of a 1.38~M$_{\sun}$ WD made of carbon and oxygen. The hydrodynamic evolution starts with the ignition of a few 
sparks and burns $\la0.3$~M$_{\sun}$, mostly to Fe-group elements, during
the first second, releasing $<5\times10^{50}$~ergs of nuclear energy. The released energy resides
for the most part in the outermost layers leading to WD pulsation. 

The details of the first phase, that spans the first second after thermal 
runaway, have been known for some time \citep{pl04,gar05,liv05} so we will only briefly sketch here 
the evolution during this phase. Even though the precise configuration at 
thermal runaway is difficult to determine \citep{hoe03}, current works suggest a multipoint ignition in which the first 
sparks are located slightly off-center \citep{gar95,woo04,wun04}. If the number of sparks is too 
small or the burning rate is too slow the nuclear energy released is not enough to unbind the star and the 
explosion fails \citep{rwh07}. This is known as the "bubble catastrophe" \citep{liv05}. The 
bubbles float to the surface before the combustion wave can propagate 
substantially \citep{pl04,gar05} and the star remains energetically bound. This behavior 
produces a composition inversion, i.e. the internal volume is plenty of fuel, cold carbon and 
oxygen, while the ashes of the initial combustion, 
mostly hot iron and nickel, are scattered around. 

The second phase of the explosion starts when the deflagration quenches due to 
expansion and ends when an accretion shock is formed (several seconds after initial thermal runaway) by the impact of the in-falling material onto the carbon-oxygen core. An important feature is that the decompression and further expansion of the cold core in his way to regain hydrostatic equilibrium imparts mechanical work to the iron-rich atmosphere. The core-atmosphere energy transfer contributes to the strength of the shock formed when the 
atmosphere re-collapses. The evolution from the time of maximum expansion up to the moment of formation of the accretion shock can be compared to simulations of the self-gravitating collapse of a cold gas sphere, a well-known hydrodynamical test that can be found, for instance in \citet{evr88}. The structures seen in the third column of their Fig.~5 qualitatively agree with those appearing in Fig.~1 of \citet{bra06} in spite of the different initial conditions.

\subsection{Numerical simulations of the pulsating phase}

We have simulated the evolution of the white dwarf during the second phase of the PRD explosion mechanism with a Smoothed Particle Hydrodynamics (SPH) code whose main features have been described elsewhere \citep{gar05}. The number of particles used in the experiments was $N=250\,000$, which provided a good spatial resolution, as detailed later. The initial model for these numerical experiments is given by the output obtained at the end of the deflagration phase computed as in \citet{gar05}. Our working hypotheses are that during the deflagration phase a mass in the range $M_\mathrm{defl}\lesssim0.30$~M$_\odot$ is burnt and that the final outcome of the explosion is determined by the value of $M_\mathrm{defl}$. We disregard any effect due to different possible deflagration histories that might lead to a given $M_\mathrm{defl}$ with different chemical compositions (e.g. different proportions of Fe-group and intermediate-mass nuclei) and chemical profiles. Hereafter, the present models will be designated with the acronym DFnn, where 'nn' stands for the hundredths of solar mass burned during the deflagration phase, for instance model DF11 means that $M_\mathrm{defl}=0.11$~M$_\odot$.

\begin{figure}
\plotone{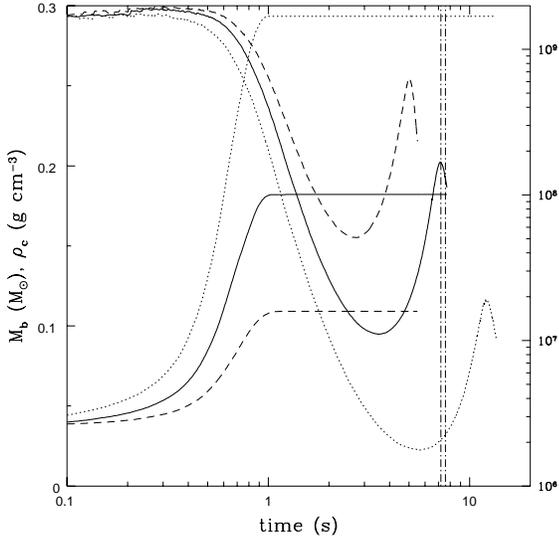}
\caption{Evolution of the models in Table~\ref{tab1} through the deflagration and pulsation phases. Shown are the evolution of the burnt mass, $M_\mathrm{b}$ (left axis scale), and the central density (right axis scale), for models DF11 (dashed lines), DF18 (solid lines), and DF29 (dotted lines). Once the white dwarf reached its minimum central density the thermonuclear burning virtually had quenched ($t\gtrsim2$~s), and the hydrodynamic evolution was thereafter followed with the same SPH code but with the nuclear reactions and the deflagration propagation algorithm switched off. The two vertical dash-dotted lines mark the times of the snapshots shown in Fig.~\ref{newfig6} (model DF18)
}\label{newfig1}
\end{figure}

In Fig.~\ref{newfig1} there is shown the evolution of models DF11, DF18 and DF29 (characterized by $M_\mathrm{defl}=0.11$~M$_\odot$, $M_\mathrm{defl}=0.18$~M$_\odot$, and $M_\mathrm{defl}=0.29$~M$_\odot$, respectively) through the deflagration and the pulsating phases. These models started from different number of igniting bubbles and used different flame velocities with the result that a range of $M_\mathrm{defl}=0.11-0.26$~M$_\odot$ was obtained. The process of generating the initial model was the same as described in \citet{gar05}, the location of the center of mass of the bubbles and their incinerated mass are given in Table~\ref{tab0}. Models DF11 and DF18 used only the first six bubbles in the Table, while model DF29 started from the seven bubbles listed. The flame velocities were: $v_\mathrm{defl} = 100$, 150, and 200~km~s$^{-1}$ for models DF11, DF18, and DF29 respectively. 

\begin{figure}
\plotone{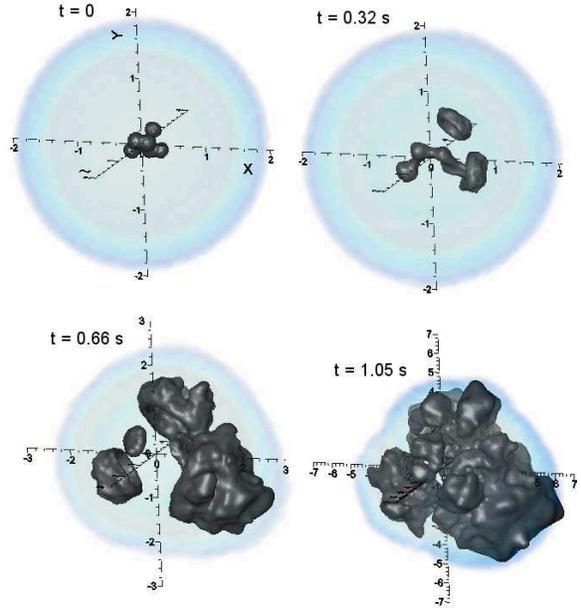}
\caption{Evolution of model DF18 during the first second of the hydrodynamical simulation (deflagration phase). Shown are the isosurfaces defined by an ash mass fraction $X_\mathrm{ash}=0.5$, and volume renderings of the density (in blue shades). The minimum density depicted is $10^6$~g~cm$^{-3}$ in the two first snapshots, $3\times10^5$~g~cm$^{-3}$ in the third snapshot, and $3\times10^4$~g~cm$^{-3}$ in the last one. The coordinate origin is at the center of mass of the white dwarf. The axes units are 1~000~km
}\label{newfig2}
\end{figure}

As can be seen in Fig.~\ref{newfig1}, most of the thermonuclear burning took place during the first second for all models. Two factors determined the quenching of the burning (in this context by quenching we mean that the nuclear timescale became much larger than the dynamical timescale): global density decrease due to white dwarf expansion, and bubbles migration to the external layers. The simulations described hereafter belong to the evolution following maximum expansion of the white dwarf. From that time on the evolution was followed with the nuclear reactions switched off in order to expose the fuel to extreme conditions during the collapse and check if a stable detonation could be achieved. Note that this is nothing but a convenient approximation to the real situation in which some preliminary burning is expected before maximum compression, thereby modifying the maximum density and temperature achieved during pulsation, as shown in paper~II. The three-dimensional evolution of the bubbles during the deflagration phase of model DF18 is shown in Fig.~\ref{newfig2}. The first noticing feature is that the evolution of the bubbles is similar to each other: all of them grow in mass through burning and rise to the surface of the white dwarf at the same time that their shape becomes toroidal due to shear forces on their top \citep{zin05,rwh07}. At $t=0.66$~s fuel takes up the whole central volume of the white dwarf. At $t=1.05$~s the flame has virtually quenched and the bubbles have expanded laterally to completely surround the core of the star. Afterwards, the ashes hide the evolution of the inner fuel during the pulsation. 

\begin{figure}
\plotone{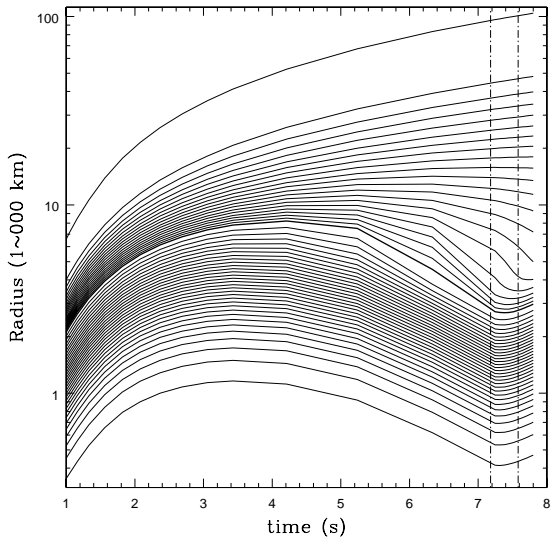}
\caption{Evolution of the white dwarf during the first pulsation of model DF18. Each curve shows the radius of a sphere that encloses a constant mass. Starting from the center outwards the mass of each curve increases by $0.0345$~M$_\odot$ up to the thick curve (which belongs to ta Lagrangian mass of $\sim1.10$~M$_\odot$), from which the increase is reduced to $0.0138$~M$_\odot$. The accretion shock forms approximately 4-5 shells above the thick line. The two vertical dash-dotted lines mark the times of the snapshots shown in Fig.~\ref{newfig6} (model DF18)
}\label{newfig3}
\end{figure}

\begin{figure}
\plotone{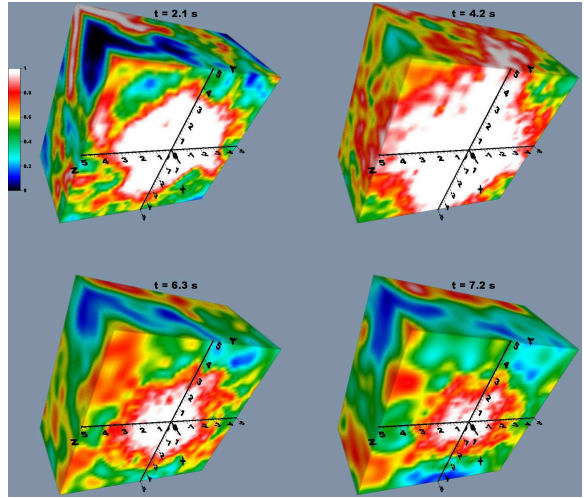}
\caption{Map of the fuel (C+O) mass fraction in and around the dense core during the first pulsation of model DF18. The three surfaces shown in the pictures belong to the planes $x=0$, $y=5\,000$~km, and $z=5\,000$~km. The axes units are 1~000~km
}\label{newfig4}
\end{figure}

\begin{figure}
\plotone{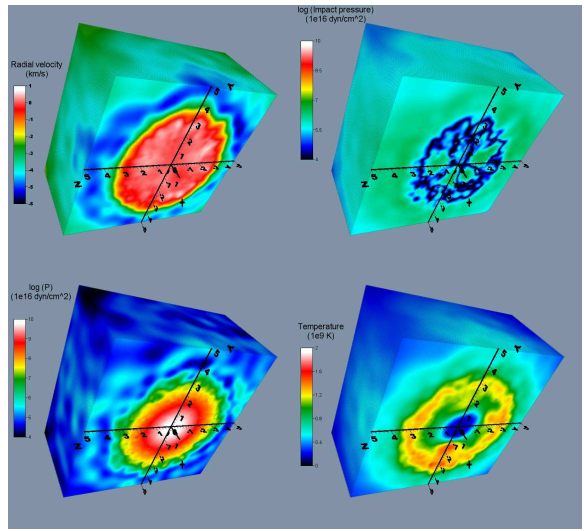}
\caption{Thermal and mechanic structure of the white dwarf at the time of accretion shock formation in model DF18 ($t\sim7.2$~s). The three surfaces shown and the axes units are as in Fig.~\ref{newfig4}. Here, as in Figs.~\ref{newfig2} and \ref{newfig4}, each three-dimensional image represents a scalar field reconstructed from the computed SPH particles properties (mass, position, velocity, smoothing length, chemical composition, temperature) through standard SPH interpolation into a regular Cartesian lattice
}\label{newfig5}
\end{figure}

The evolution of model DF18 during the first pulsation, after deflagration quenching, is shown in Figs.~\ref{newfig3} and \ref{newfig4}, and the structure of the white dwarf at the time of accretion shock formation, $t_\mathrm{acc}$, is shown in Fig.~\ref{newfig5}. The inner $\sim1.10$~M$_\odot$ pulsate in phase and reach a state of maximum compression at $t=7.16$~s, at which time they make up a compact nearly hydrostatic core. At the same time, the outermost layers continue expanding with velocities larger than the escape velocity, while the intermediate layers continue falling onto the core with velocities as large as $5\,000$~km~s$^{-1}$. The successive stretching and shrinking of the C-O rich core can be better seen in Fig.~\ref{newfig4}. Although the chemical structure carry the imprint of the bubbles evolution and lacks spherical symmetry, but for the central volume, the fuel mass fraction at radius below $\sim5\,000$~km is high: $X_\mathrm{fuel}\gtrsim0.4$. 

The mechanical and thermal structure at $t_\mathrm{acc}$ display a high degree of spherical symmetry, as shown by Fig.~\ref{newfig5}. The picture of the radial velocity reveals a large gradient above $r\sim3\,000$~km. The opposed gradients of radial velocity and density lead to a quite homogeneous impact pressure (due to matter infall onto the core) as high as $P_\mathrm{ram}=\rho v^2\simeq10^{23}$~dyn~cm$^{-2}$ in the range of distances $r\sim3\,000-5\,000$~km. This impact pressure takes over the thermal pressure above $r\sim4\,000$~km, so that the largest deceleration and subsequent heating of the in-falling matter takes place in a ring (in fact, a narrow shell in three dimensions) at $r\sim3\,000-4\,000$~km, as seen in the temperature plot. The fuel mass fraction in these locations at $t_\mathrm{acc}$ lies in the range $X_\mathrm{fuel}\sim0.4-0.8$.

\subsection{Comparison between PRD and PDD models}\label{PDD}

In order to fully understand the implications of the three-dimensional 
calculations of the pulsating phase of the supernova it is convenient to 
compare the results of the present PRD simulations with those obtained with a one-dimensional code (PDD models).
The key difference between the simulations performed in one and three dimensions is the 
impossibility of the former to change the sequence of the mass elements. This has two important consequences. First, in one-dimensional
calculations the material burned close to the center never catches the fuel located ahead of it. Second, in one-dimensional calculations the unburned mass always lies above the flame blocking the expansion of the underlying layers and facilitating a more efficient burning. In contrast, in three-dimensional models the 
floatability of igniting bubbles allows them to reach the external layers with 
only a moderate transfer of mechanical work to their surroundings. Thus, in three dimensional PRD calculations the 
ashes retain most of their specific energy (kinetic and thermal), which enables the formation of a robust accretion shock that  ultimately triggers a detonation. In contrast, in the PDD mechanism, the detonation is thought to be due to compression of the mixing layer between unprocessed and processed material, which is believed to grow only if there is a long-range pulsation of the white dwarf \citep{hoe95}.

\begin{figure}
\plotone{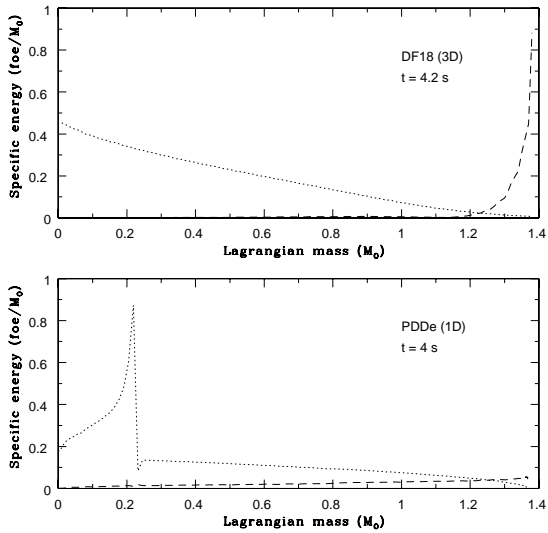}
\caption{Distribution of specific energy within the white dwarf during the 
first pulsation in a one-dimensional calculation of a pulsating delayed detonation model (model PDDe in Badenes \etal\ 2003) vs the three-dimensional model DF18: specific thermal energy (dotted-line) and kinetic energy (dashed-line). The top panel shows profiles obtained from angle-averaged versions of the 
DF18 model at a time t=4.2~s after initial runaway. The bottom panel shows the profiles of model PDDe at a similar elapsed time
}\label{fig7}
\end{figure}

\begin{figure}
\plotone{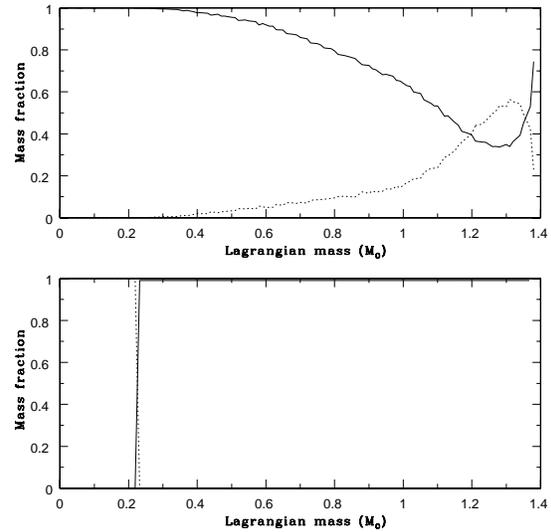}
\caption{Same as previous Fig., but for the chemical composition. 
Solid-line: C+O, dotted-line: products of the incineration to nuclear 
statistical equilibrium (mainly Fe-Ni)
}\label{fig8}
\end{figure}

In Figs.~\ref{fig7} and \ref{fig8} there are represented the angle-averaged specific energy 
and chemical profiles, respectively, of the  
three-dimensional model DF18 (top panel) and a one-dimensional calculation of a 
pulsating delayed detonation model \citep[bottom panel: model PDDe in][]{bad03}. In the three-dimensional calculations the energy  
resides for the most part in the outer 0.15~M$_{\sun}$. Hence, the pulsating 
motion of this material is decoupled from the rest of the structure. In 
contrast, in the one-dimensional calculation the energy is concentrated in the 
inner 0.2~M$_{\sun}$, that behaves like a piston driving the pulsation 
of most of the structure nearly in phase. In the PDD mechanism the detonation is initiated at the edge of the incinerated core during the collapsing phase and propagates outwards through the in-falling matter \citep{kho93}. 

An important requirement of the PDD mechanism is the necessity to burn a quite precise mass of fuel, $\sim0.3$~M$_\odot$, 
during the initial deflagration phase in order to achieve a large amplitude 
pulsation (necessary to create a wide enough mixed layer) yet not unbind the white dwarf. This is not the case in the PRD 
models, because:
\begin{enumerate}
\item The detonation is not launched as a result of turbulent mixing of fuel 
and ashes, thus avoiding the necessity to achieve a large amplitude of the 
pulsating motion.
\item The relative amplitude of the pulsation is not uniform for all the mass 
layers of the white dwarf. The external layers experience a large amplitude 
because they are rich in hot ashes, freshly synthesized in the floating burning 
bubbles, while the 1~M$_{\sun}$ core remains at moderately high density during the pulsation.
Thus, in the PRD scenario a large amplitude of the external layers is 
obtained irrespectively of the precise amount of mass burnt during the deflagration 
phase.
\end{enumerate}

\begin{figure}
\plotone{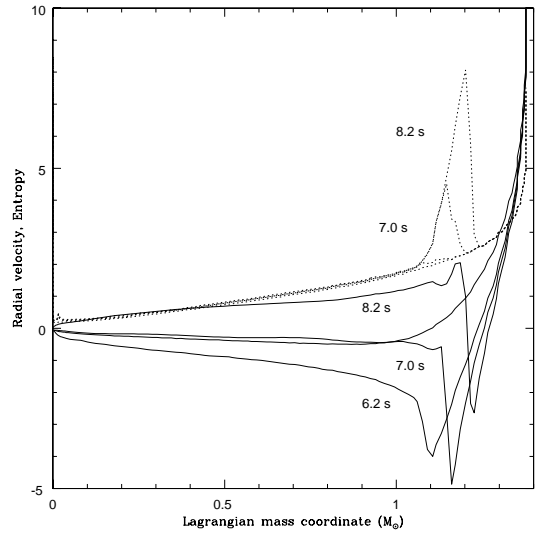}
\caption{Evolution of velocity (solid lines, in 1,000 km~s$^{-1}$) and entropy (dotted lines, in $k_\mathrm{B}$ per particle) calculated with a one-dimensional code after an angle-averaged version of the DF18 model starting at $t=4.2$~s. Times plotted are 4.2~s, 6.2~s, 7.0~s, and 8.2~s. For clarity, neither the density profile at 4.2~s nor the entropy profiles at 4.2~s and 6.2~s have been labelled
}\label{figPRD}
\end{figure}

\begin{figure}
\plotone{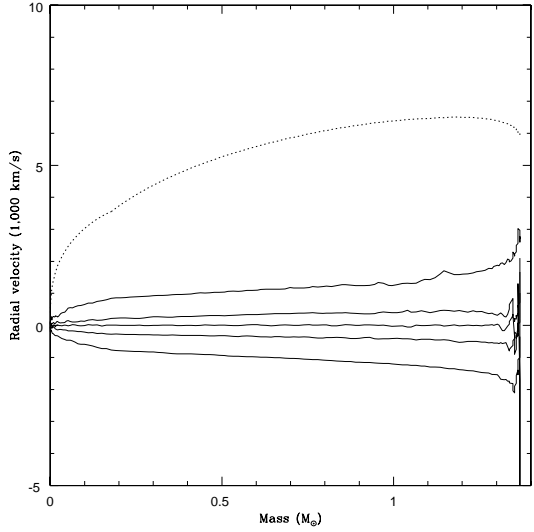}
\caption{Evolution of velocity of a PDD model that burns subsonically the same mass as model DF18. Times plotted are 4.2~s, 6.2~s, 7.0~s, 8.2~s, and 11.4~s. The dotted line gives the escape velocity at $t=4.2$~s
}\label{figPDD}
\end{figure}

Finally, we compare the evolution during the pulsation of the structures obtained at the end of the deflagration phase for the three-dimensional model DF18 and that of a one-dimensional model constrained to burn subsonically the same mass as DF18. First, we have mapped the DF18 model at 4.2~s into an angle-averaged version and its posterior evolution has been followed with a one-dimensional hydrocode\footnote{The one-dimensional hydrocode, based on the finite differences scheme proposed by \citet{cw66}, is the one described in \citet{bad03} and \citet{bra93}} with the nuclear reactions turned off. A sequence of the velocity and entropy profiles obtained this way is shown in Fig.~\ref{figPRD}. In this Fig. it can be appreciated clearly the process of formation of the accretion shock, whose main distinctive feature is the sudden increase of the specific entropy. Second, we have computed a one-dimensional PDD model with $M_\mathrm{defl} = 0.18$~M$_\odot$. By construction, the difference between this PDD model and the one-dimensional version of DF18 resides in the distribution of fuel and ashes, as well as the specific energy they transport. The consequences are plain to see in Fig.~\ref{figPDD}, in which the evolution of the velocity profiles of the PDD model is displayed. Except for a few layers at the top of the white dwarf, the star pulsates in phase without development of any shock wave (unless a detonation is assumed to happen). 
These calculations show that the chemical composition inversion is a crucial ingredient in the evolution of the white dwarf up to the moment of detonation ignition.

\section{Confined detonation ignition}

We now turn to the question of the robustness of the CDI in the PRD scenario. First we examine the physical conditions in the accretion shock, as obtained in three-dimensional simulations of the pulsating phase. By comparing these conditions with suitable detonation criteria, we then analyze the most favorable conditions for detonation ignition in our three-dimensional simulations.

\subsection{Physical conditions of shocked matter}\label{phys}

The location and physical conditions at the accretion shock are very sensitive to the amount of mass burned during the initial deflagration phase. A summary of the properties of the accretion shock for three calculated models is given in Table~\ref{tab1}. The mass at which the accretion shock is formed, $M_\mathrm{acc}$, is larger for smaller $M_\mathrm{defl}$ because less energy is available to power the expansion of the white dwarf. The relationship between both masses can be approximated in the range of $M_\mathrm{defl}$ explored by: $M_\mathrm{acc}\approx-4.18M_\mathrm{defl}^2-1.22M_\mathrm{defl}+1.50$, where all masses are expressed in M$_\odot$. Other quantities that can be fitted to an analytic function of $M_\mathrm{acc}$ are: the time of formation of the accretion shock, $t_\mathrm{acc}\approx-15.9M_\mathrm{acc}+25.1$~s, and the central density at that time, $\rho_\mathrm{c8}\approx\exp\left(8.87M_\mathrm{acc}^2-11.0M_\mathrm{acc}+1.50\right)$, where $\rho_\mathrm{c8}$ is in units of $10^8$~g~cm$^{-3}$. 

Once formed, the accretion shock remains confined close to the hydrostatic core due 
to the large ratio of the impact pressure of the in-falling matter with respect to the gas 
pressure: $\rho v^2/p = \gamma M^2 \approx 9-12$, where $\rho v^2$ is the impact 
pressure, $p$ is the gas pressure, $\gamma=5/3$ is the adiabatic coefficient, 
and $M$ is the Mach number of the flow. The density at the shock is low, $\rho_\mathrm{shock}\approx(1.4-5.2)\times10^5$~g~cm$^{-3}$, and the temperature is high, 
\begin{equation}
T_\mathrm{shock}=\frac{3}{16} \frac{\mu}{kN_\mathrm{A}} v^2 \approx (0.5-1)\times10^9~\mathrm{K}\,,
\end{equation}
\noindent where $\mu$ is the mean molar mass ($\mu =1.75$~g~mol$^{-1}$ for completely ionized carbon and oxygen matter), $N_\mathrm{A}$ is Avogadro's number, and $k$ is the Boltzmann constant. The accretion shock evolution is thereafter driven by two opposite effects: inertial confinement due to mechanical energy deposition by in-falling matter vs pressure build-up due to nuclear energy release (not included in the present simulations). Eventually, the nuclear energy released by the burning of a mass $M_\mathrm{pri}$ becomes large enough for pressure build-up to take over the mechanical energy deposition, causing the accretion shock to expand and decouple from the underlying dense hydrostatic core. If this {\sl primer} mass is large enough to ignite a detonation, then nuclear processing of most of the core will be ensured, otherwise nuclear burning will cease shortly after due to adiabatic cooling of the core. 

The rate of mechanical energy deposition at the accretion shock, $\varepsilon_\mathrm{mech}=2\pi r^2 \rho_0 v^3$, stays in the range $(0.5-1)\times10^{49}$~erg~s$^{-1}$ during approximately $\Delta t\sim 0.3$~s. Assuming that nuclear burning proceeds just to $^{28}$Si (because $\rho<10^7$~g~cm$^{-3}$) the specific nuclear binding energy released is $q\sim5.8\times10^{17}$~erg~g$^{-1}$. An estimate of the maximum primer mass that can be accumulated before expansion of the accretion shock can be obtained by equating the nuclear energy release to the mechanical energy deposited by in-falling matter $M_\mathrm{pri} q \approx\varepsilon_\mathrm{mech} \Delta t$, which gives $M_\mathrm{pri}\approx(2.6-5.2)\times10^{30}$~g. 

The radius at which a detonation could ignite can be estimated independently from the above calculations in the following way. 
A necessary condition for a detonation to be initiated is that the nuclear timescale must be lower than the hydrodynamical timescale, $\tau_\mathrm{hydro} = 446/\sqrt{\rho_\mathrm{shock}} \approx 0.6-1.2$~s. The nuclear timescale of carbon burning, $\tau_\mathrm{nuc}$, can be calculated as \citep{bk87,kho89}:
\begin{equation}
\tau_\mathrm{nuc} = \Theta\tau_\mathrm{i}\,,
\end{equation}
\noindent where $\Theta$ is the Frank-Kamenetskii factor, 
\begin{equation}
\Theta = \left(\frac{QT_{9A}^{2/3}}{3T_9}\right)^{-1}\,,
\end{equation}
\begin{equation}
T_{9A} = T_{9}/\left(1+0.067T_9\right)\,,
\end{equation}
\noindent $Q = 84.165$, and $T_9 = T /10^9$~K. Finally, $\tau_\mathrm{i}$ is obtained as:
\begin{equation}
\tau_\mathrm{i} = \frac{c_\mathrm{e}T}{\rho q_\mathrm{c} A_{T9} Y_\mathrm{C}^2
\exp\left(Q/T_{9A}^{1/3}\right)}\,,
\end{equation}
\noindent where $q_\mathrm{c} = 4.48\times10^{18}$~erg~mol$^{-1}$, $c_\mathrm{e}$ is the specific heat, $A_{T9} = 8.54\times10^{26}T_{9A}^{5/6} T_9^{-3/2}$, and $Y_\mathrm{C} = 0.5/12$~mol~g$^{-1}$ is the $^{12}$C molar fraction (we assume a 50\%-50\% by mass carbon-oxygen composition).

\begin{figure}
\plotone{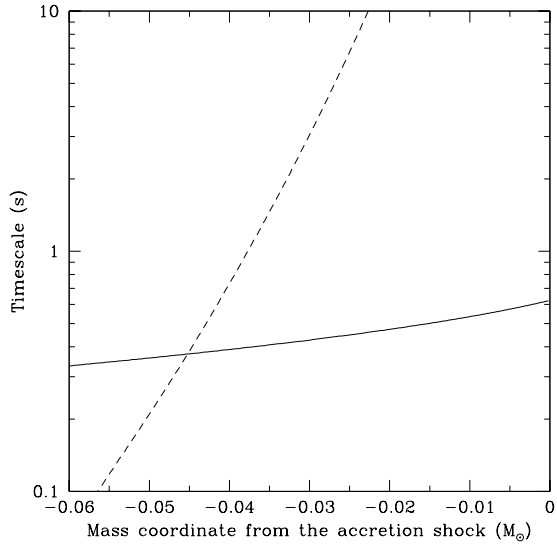}
\caption{Comparison of the hydrodynamic timescale (solid line) and the timescale of carbon 
burning (dashed-line) within the shocked flow. The zero of the horizontal axis marks the location of the accretion shock
}\label{fig9}
\end{figure}

Using the above equations with the values of $\rho_\mathrm{shock}$ and $T_\mathrm{shock}$ from Table~\ref{tab1} the nuclear timescale is larger than the hydrodynamical timescale, hence a detonation cannot begin just behind the accretion shock. However, the condition $\tau_\mathrm{nuc} < \tau_\mathrm{hydro}$ can be met during the journey of the shocked matter along its path towards the surface of the hydrostatic core. 
In order to obtain an analytical estimate of the distance from the accretion shock at which the timescale requirement can be fulfilled we need to make some simplifying assumptions concerning the shocked flow: 
\begin{itemize}
\item Steady state during times small compared with $\tau_\mathrm{hydro}$.
\item Spherical symmetry and radial flow (we ignore any instability in the flow).
\item Optically thick matter (we assume thermal equilibrium with radiation).
\item Adiabatic evolution (we ignore the nuclear energy input and any energy transfer mechanism, plus optically thick matter which implies that radiative cooling is inefficient).
\end{itemize}
Under these assumptions, the structure of the shocked flow in between the 
accretion shock and the core surface can be obtained by solving the following 
set of equations:
\begin{equation}\label{eq8}
e = \frac{1}{2}v^2 + \frac{p}{(\gamma-1)\rho} - \frac{GM}{r} =
\mathrm{constant}\,,
\end{equation}
\begin{equation}\label{eq9}
\rho v r^2 = \mathrm{constant}\,,
\end{equation}
\begin{equation}\label{eq10}
p \propto \rho^\gamma\,,
\end{equation}				
\noindent The starting point for the integration of these equations is just the physical state behind the shock (see Table~\ref{tab1}):
$\rho_\mathrm{shock}$, and $T_\mathrm{shock}$.
The hydrodynamical and nuclear timescales obtained from the solution to the above equations for model DF18 are plotted in Fig.~\ref{fig9}, where it can be seen that carbon burns in less than a hydrodynamical time at a radius of $\sim3\,000$~km. The density and temperature at that point are given by the cross symbol in Fig.~\ref{detcond}.

As shown in Table~\ref{tab1}, the distance from the accretion shock to the detonation ignition point is sensitive to the mass burned subsonically. The Lagrangian mass at which a detonation could begin, $M_\mathrm{deto}$, goes from 1.03~M$_\odot$ for model DF26 to 1.29~M$_\odot$ for model DF11, which is 0.06~M$_\odot$ to 0.03~M$_\odot$ inwards from the accretion shock, respectively. The value of $M_\mathrm{deto}$ quoted in Table~\ref{tab1} for model DF29, $M_\mathrm{deto}=0.38$~M$_\odot$ is far too low to be minimally realistic.
However, we stress that the analytic calculations of the formation of a detonation 
presented in this section are just approximations because of the many simplifications introduced, particularly 
the assumption of spherical symmetry and steady state. In the three-dimensional picture there are to be expected some deviations from the results of these analytic calculations.

\subsection{Detonation ignition conditions}

Small-scale simulations are needed to determine which are the suitable conditions for detonation ignition in white dwarf matter. 
\citet{arn94,nie97,rwh07} performed such studies following the hydrodynamical and nuclear evolution of a uniform density microsphere. This thermonuclear bomb was lit using a primer consisting in a thermal profile characterized by a central peak temperature and a constant negative thermal gradient down to $10^8$~K at a Lagrangian mass coordinate $M_\mathrm{pri}$, beyond which the temperature was uniform. Once ignited, the sphere was subject to two opposite effects: it expanded and cooled due to the excess of pressure derived from the sudden incineration of the hot primer but, at the same time, the released nuclear energy helped keeping a high value of temperature. Which one of the two effects wins is what determines if a detonation is successfully launched or not. These numerical experiments are therefore characterized by three parameters: the central temperature, $T_\mathrm{c}$, the uniform sphere density, $\rho$, and the mass of the primer, $M_\mathrm{pri}$. A larger primer mass implies a larger energy reservoir and a stronger piston effect, making it easier to build a detonation wave. The detonation initiation conditions in pure C-O matter (50\% each by mass) obtained by \citet{arn94,nie97,rwh07} are summaryzed in Table~\ref{tab2}. Their results can be interpreted either as the minimum primer mass for which a detonation is initiated in C-O matter at a given density and peak temperature, or as the minimum temperature for which a detonation is obtained at a given density and mass of the primer. In any case, the qualitative trend is that the smaller the density the harder is to initiate a detonation while, at fixed density, the higher the peak temperature the lesser primer mass is needed to detonate the sphere. From Table~\ref{tab2} it stems that at densities in the range $10^7 - 10^8$~g~cm$^{-3}$ a temperature of $\sim2\times10^9$~K is enough to initiate a detonation. At lower densities the primer mass necessary to initiate a detonation at $2\times10^9$~K becomes a non-negligible fraction of the star mass, hence achievement of detonation initiation conditions is much more difficult. At $\rho = 3\times10^6$~g~cm$^{-3}$ and $T = 2.3\times10^9$~K a detonation was obtained for $M_\mathrm{pri} = 2\times10^{28}$~g, implying a thermal gradient $\nabla T = 200$~K~cm$^{-1}$, of the same order of the maximum thermal gradient estimated by using Eq.~\ref{eq1}:  $\nabla T_\mathrm{max} = 5 - 150$~K~cm$^{-1}$. 

\begin{figure}
\plotone{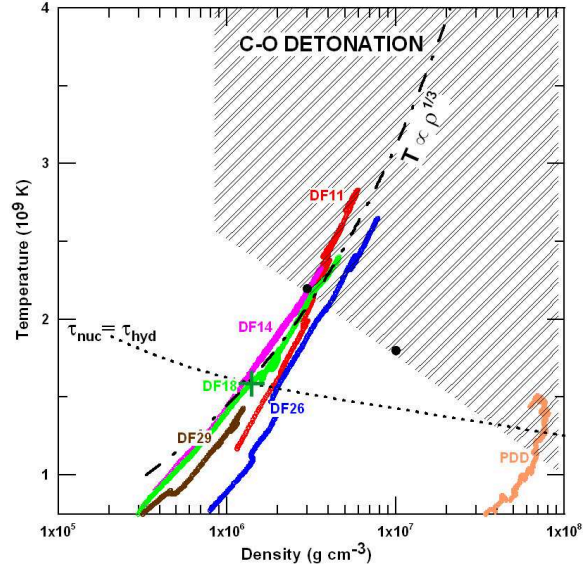}
\caption{Maximum temperature and density reached during the pulsating phase
of several models. The colored points show the track of the hottest shocked fuel
particles of each model. The shaded area is the region where a C-O
detonation can be successfully launched, extrapolated from the two first rows in Table~\ref{tab2}
\citep{rop07}. The dash-dotted line represents a $T\propto\rho^{1/3}$ relation, plotted for reference. The dotted line shows the location of $\tau_\mathrm{nuc} = \tau_\mathrm{hydro}$. The point at which this line is crossed in the DF18 model according to the solution of Eqs.~\ref{eq8}-\ref{eq10} is given by the dark-green cross symbol. The time at which different models achieve detonation ignition conditions (at the entrance to the shaded area) is given in the last row of Table~\ref{tab1}. The SPH models summarized in this figure did not include nuclear reactions
}\label{detcond}
\end{figure}

The results of the small-scale simulations allow us to look for regions in the three-dimensional hydrodynamic simulations that meet some gross criteria for detonation ignition.
A comparison of the results of the present three-dimensional simulations of white dwarf pulsation with the detonation ignition conditions summarized in Table~\ref{tab2} is shown in Fig.~\ref{detcond} where we plot in the $\rho$-$T$ plane the evolution of the fuel particles that achieve the most favorable conditions for detonation ignition, i.e. those particles that attain the maximum temperature after being shocked. Shown in this Fig. are the results for five models for which $M_\mathrm{defl}$ spans the range from 0.11~M$_\odot$ to 0.29~M$_\odot$. The two big dots belong to the two first rows in Table~\ref{tab2}, which are used to delimit the $\rho$-$T$ conditions for which a C-O detonation is the most probable outcome. As can be seen in this Fig., such conditions are clearly reached in all the three-dimensional pulsating models except for DF29, characterized by a mass burned subsonically close to the mass needed to unbind the white dwarf (0.30~M$_\odot$). The path followed by the particles belonging to the different models is nearly identical, and very close to adiabatic evolution (dot-dashed line in Fig.~\ref{detcond}). The point at which model DF18 (green points) crosses the line defined by $\tau_\mathrm{nuc} = \tau_\mathrm{hydro}$ is in good agreement with the estimate based on Eqs.~\ref{eq8}-\ref{eq10} (dark-green cross). In all cases but model DF29, the C-O detonation ignition line is crossed at $\sim3\times10^6$~g~cm$^{-3}$. The peaks of temperature and density span a narrow range: $T_\mathrm{max}\approx (2.4-2.9)\times10^9$~K and $\rho_\mathrm{max}\approx (4-8)\times10^6$~g~cm$^{-3}$, high enough to confidently assume that a CDI will be produced in these cases. 

\begin{figure}
\plotone{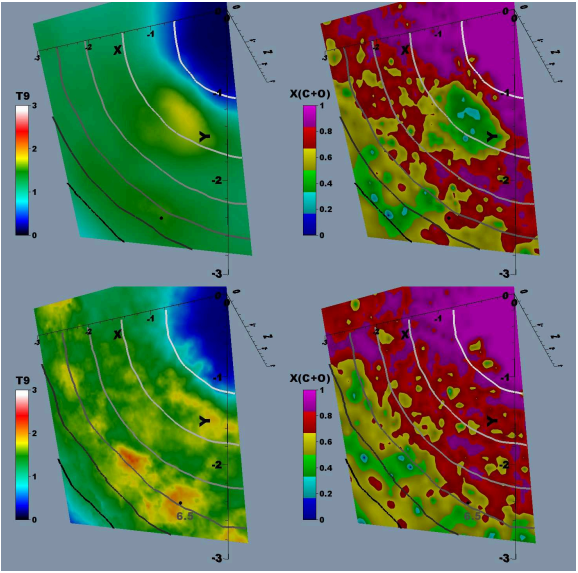}
\caption{Thermodynamic structure (left) and fuel mass fraction (right) in a slice of model DF18 in the neighborhood of the hottest shocked fuel particle plotted in Fig.~\ref{detcond}. The top row belongs to a time shortly before the formation of the accretion shock, while the bottom row belongs to the time at which the particle reaches C-O detonation conditions (shaded area in Fig.~\ref{detcond}), i.e. $\sim0.3$~s later than the top row (see Table~\ref{tab1}, and Figs.~\ref{newfig1} and \ref{newfig3}). The position of the particle is shown by a black dot located close to the bottom and to the center of each slice. The temperature is given in units of $10^9$~K, while the density contours belong to, from the center outwards and in units of g~cm$^{-3}$, $\mathrm{log}\rho=8$, 7.5, 7, 6.5, 6, and 5.5 (the contour belonging to $\mathrm{log}\rho=6.5$ is labelled in the bottom row slices for clarity). The images have been obtained from the computed properties of the SPH particles as explained in Fig.~\ref{newfig5}. The axes units are 1~000~km and their directions are the same as in Figs.~\ref{newfig2}, \ref{newfig4}, and \ref{newfig5}
}\label{newfig6}
\end{figure}

In Fig.~\ref{newfig6} there are shown the thermodynamic and chemical structure in a slice of model DF18 in the neighborhood of the hottest shocked fuel particle before (top row) and after (bottom row) it reaches detonation ignition conditions ($T\sim2.3\times10^9$~K, $\rho\sim3\times10^6$~g~cm$^{-3}$). The detonation conditions are first reached in a hot spot of radius $\sim200$~km (mass $\sim10^{29}$~g) affecting a region with non-uniform chemical composition, where the fuel mass fraction is in the range $X(\mathrm{C+O})\sim0.4-0.8$. A rough estimate of the thermal gradient across the hot spot gives $\nabla T \sim 10^9~\mathrm{K}/200~\mathrm{km} = 50$~K~cm$^{-1}$. Thus, although the situation is much more complex in our models than that addressed in the small-scale simulations, we can confidently expect that a stable detonation will be obtained a few tenths of a second after the formation of the accretion shock.
Note that the hot spot that appears close to the center in the top row image is due to the presence of a clump of ashes, which avoids a detonation to be initiated at its position. 
Note also that, because we did not allow for nuclear reactions in the present models once the deflagration phase ended, the chemical changes between both snapshots are only due to advection of particles of different composition.

The maximum temperature and density reached in model DF29, $T_\mathrm{max}=1.42\times10^9$~K and $\rho_\mathrm{max}=1.3\times10^6$~g~cm$^{-3}$, are seemingly too low for igniting a stable detonation. \citet{rwh07} did not obtain a stable detonation for $\rho=10^6$~g~cm$^{-3}$ and a temperature of $T=3.0\times10^9$~K even for a primer mass as high as $M_\mathrm{pri}=3.0\times10^{30}$~g. This mass is comparable to the maximum primer mass derived in Sect.~\ref{phys} for model DF29: $M_\mathrm{pri}=2.6\times10^{30}$~g (see also Table~\ref{tab1}).

For comparison purposes we show in Fig.~\ref{detcond} the path followed during the recontraction phase by the PDD model described in Section~\ref{PDD}, but with the nuclear reactions turned off. The displayed path is that drawn by a fuel particle in the neighborhood of the quenched flame. The path of the particle penetrates the C-O detonation region at $\rho\sim7\times10^7$~g~cm$^{-3}$ and $T\sim1.2\times10^9$~K, and afterwards reaches a maximum temperature of $T_\mathrm{max}=1.5\times10^9$~K. However, note that this region of the $\rho$-$T$ plane has not been thoroughly explored in order to determine the conditions suitable for detonation ignition (see Table~\ref{tab2} and references therein). Thus, in this region the threshold line is just an extrapolation from lower density points. It is therefore clear that the conditions achieved during the re-contracting phase of PDD models are quite different from those obtained with PRD models. 

\subsection{Resolution issues}

Our numerical resolution is much coarser than the expected detonation width at the densities of detonation ignition. Thus, we have explored the sensitivity of the maximum temperature and density reached during white dwarf pulsation to the resolution of the numerical models. To this end we have computed additional simulations of the pulsating phase of models DF18 and DF29 with increased resolution. The starting point of these additional models was the time of maximum expansion during the pulsation. The particles that fulfilled certain selection criteria were splitted into $N_\mathrm{sp}$ children particles, and the star evolution was again followed with the same three-dimensional SPH code described before (the term {\it particle splitting} was used for the first time by \citealt{kit02}). In order to save computer time, splitting was confined to the neighbors of the shocked fuel particles that reached the maximum temperature (Fig.~\ref{detcond}), and to the neighbors of their neighbors. We performed simulations for $N_\mathrm{sp}$=2, 4, 10, and 100. In every case the original particle mass was shared evenly between their children, which resulted in a maximum mass resolution of $1.1\times10^{26}$~g. The initial position of the children particles was randomly selected within a sphere centered on the position of the parent particle and radius given by its smoothing length. They were therefore assigned the same temperature and chemical composition as the parent particle, while their velocity was obtained by interpolation of the velocity field at the position of the children. 

The results of our resolution study are shown in Table~\ref{tab3}, where there are given the maximum density and temperature reached by children particles as well as their maximum spatial resolution, $\Delta x_\mathrm{min}$. Even though the maximum temperature and density for a given $M_\mathrm{defl}$ show some scatter with increasing resolution they all stay at the same side of the C-O detonation line. The departure from the conditions obtained with our standard non-splitted models is quite modest and does not affect our conclusions: all the variants of model DF18 fulfilled the conditions for a C-O detonation, while all the variants of model DF29 did not. 

In our highest resolution simulations of model DF18, the increase in resolution does not lead to a too steep thermal gradient that might compromise the achievement of detonation conditions. In model DF18sp100, the primer mass (calculated as the total mass of fuel particles with $\rho \geq 3\times10^6$~g~cm$^{-3}$ and $\nabla T \leq 200$~K~cm$^{-1}$ that are neighbors of the hottest children) rises to $1.8\times10^{28}$~g. This mass is similar to the minimum mass required to detonate C-O, as given in Table~\ref{tab2} for the same density but for a peak temperature substantially lower than the maximum temperature achieved in model DF18sp100. As we have not been able to achieve convergence of the maximum temperature and density, it is not clear if further increase of the resolution might result in a thermal gradient too steep to launch a detonation. 

\section{A helium cap on the white dwarf?}

One way to favor a detonation ignition in C-O matter is to mix it with a small amount of He, which both is  more reactive and releases more nuclear energy than carbon. To allow such a mixing, some mass of He must be present in the white dwarf at the moment of thermonuclear runaway. Indeed, a very small cap made of He might rest atop of the white dwarf as a result of previous accretion from the companion star in the binary system. This helium could either have been accreted directly from a He star or result from the nuclear processing of accreted hydrogen. \citet{hk96} suggested that the presence of a $\sim0.01$~M$_\odot$ cap of He would improve the match between calculated and observed light curves. This amount of He might come from the surrounding accretion disk if the accretion rate were not constant, for instance if it dropped at some time below the critical rate for steady burning. \citet{cu94a,cu94b} detected lines coincident with two He~I lines at 2.04~$\mu$m and 1.052~$\mu$m in the spectra of the spectroscopically normal SN 1994D. The absorption feature at 1.05~$\mu$m has been since detected in the early spectra of many other SNIa \citep[see][]{no03,pi08}, although its identification is problematic because that line could rather be due to Mg~II \citep{wh98,ha99}. However, as pointed out by \citet{ml98}, He~I lines may be blended with magnesium lines as well as with lines of other intermediate-mass elements. Hence, up to $\sim0.01$~M$_\odot$ of He might remain undetected in SNIa, hidden by stronger magnesium lines.

If there was actually such small cap of helium, it would have several chances of mixing with the C-O outer layers of the  white dwarf. The first opportunity would arise during the convective simmering phase shortly before the hydrodynamic event \citep{kuh06,pb08,ch08}. Even though convection during simmering is not expected to reach the white dwarf surface, it can excite pulsating and non-spherical mode instabilities of the star that might allow diffusion of He into the underlying C-O layers. A second chance for mixing He would take place during the deflagration phase of the explosion itself. Turbulence and, especially, Rayleigh-Taylor instability induce large radial excursions of huge volumes of ashes that destabilize the whole white dwarf, and can drive the mixing of an eventual external He layer with the C-O beneath it. Hence, even though it may seem a speculative scenario, one cannot discard the possibility that He is present in some small mass fraction within the matter that is prone to detonate after a white dwarf pulsation. Mixing of He within a detonating volume seems more likely in scenarios in which the transition to detonation develops in the outermost layers of the white dwarf, e.g. those CDI described in \citet{pl04}, \citet{bra06} and \citet{rwh07}.

We have explored the conditions for stable detonation ignition in white dwarf matter composed by C-O contaminated with He, following a methodology similar to that of \citet{arn94} and \citet{nie97}. Using a one-dimensional code we have followed the hydrodynamical and nuclear evolution of a uniform density sphere made of equal amounts of carbon and oxygen with a small fraction of He. The setup of these numerical experiments, as well as the physics included, were similar to those described in \citet{nie97}. These numerical experiments are characterized by four parameters: the central temperature, $T_\mathrm{c}$, the uniform sphere density, $\rho$, the mass of the primer, $M_\mathrm{pri}$, and the He mass fraction, $X(\mathrm{He})$. In order to limit the dimensionality of the problem we have performed calculations at a fixed density: $\rho=1.5\times10^6$~g~cm$^{-3}$, while we have explored the following ranges of the rest of parameters: $1.0\times10^9~\mathrm{K} \leqslant T_\mathrm{c} \leqslant 1.8\times10^9~\mathrm{K}$, $3\times10^{25}~\mathrm{g} \leqslant M_\mathrm{pri} \leqslant 3\times10^{29}~\mathrm{g}$, $0.02 \leqslant X(\mathrm{He}) \leqslant 0.20$. Besides, two different He distributions have been explored: either it was concentrated in the primer volume, or it was uniformly distributed throughout the whole microsphere. For the extreme values of primer mass considered the ball sound crossing time ranges from 0.01~s ($M_\mathrm{pri} = 3\times10^{25}$~g) to 0.3~s ($M_\mathrm{pri} = 3\times10^{29}$~g). 

Our results, summarized in Table~\ref{tab4}, show that addition of a small quantity of He makes detonation initiation far easier than in pure C-O. In the numerical experiments in which He was constrained to the primer its role was limited to provide a larger nuclear energy release as compared to pure C-O, while distributing it uniformly throughout the sphere allowed also to take advantage of the larger reactivity of He. Table~\ref{tab4} shows that with a He mass fraction of 10\% a detonation is feasible even at a peak temperature as small as $10^9$~K provided that a large enough primer mass ($>1.5\times10^{-4}$~M$_\odot$) is present. Note that in this case the total mass of He needed to detonate the microsphere is just $>1.5\times10^{-5}$~M$_\odot$. For a 5\% He mass fraction the minimum temperature needed to achieve a stable detonation rises to $1.8\times10^9$~K for the same primer mass. The calculations with a peak temperature of $1.4\times10^9$~K and $M_\mathrm{pri}=3\times10^{27}$~g show interesting results: a 20\% mass fraction of He constrained to the primer was not able to initiate a detonation, while a modest 10\% He mass fraction uniformly distributed was enough to detonate the microsphere. These results suggest that the main reason by which He facilitates detonation ignition in C-O matter is because of its large reactivity rather than because of its large nuclear energy release. 

The application of the present results to the question of the feasibility of a CDI in model DF29 is inconclusive. Even disregarding the speculative nature of the presence of a He cap onto the white dwarf, the small Lagrangian mass at which the accretion shock forms in model DF29, $\sim0.38$~M$_\odot$, poses another difficulty to mixing He down to such a deep place. In spite of these uncertainties, we have simulated the detonation phase of model DF29 contaminated with He, whose results will be presented in detail in paper~II: for $M_\mathrm{defl}=0.29$~M$_\odot$ we have not been able to obtain a stable detonation irrespectively of the presence of any reasonable amount of He on top of the white dwarf.

\section{Conclusions}

The initiation of a detonation is a highly non-linear process in which a modest variation in the environmental conditions can have a large impact on the outcome. Theoretical analysis, small-scale simulations and large-scale numerical simulations of white dwarf explosions powered by slow subsonic flames do not favor presently a deflagration-to-detonation transition during the expansion phase of the explosion, although it is not discarded either. The exception may be DDT induction by transition of the flame to the distributed combustion regime \citep{rop07,woo07b}, but there remain many uncertainties about its feasibility in an exploding white dwarf\citep{lis00,pan08}. 

We have shown that a confined detonation ignition can be induced by an accretion shock, formed around a carbon-oxygen core after WD pulsation following subsonic burning of a mass $M_\mathrm{defl}\lesssim0.3$~M$_\odot$. The mechanism of CDI is robust, and detonation ignition conditions are reached non-marginally within the shocked flow for a wide range of initial configurations. However, if the nuclear energy released before the pulsation is close to the white dwarf binding energy the accretion shock is not efficient enough to detonate the mixture of carbon and oxygen. 

We have also addressed the effects that the presence of a He cap on top of the white dwarf, at the time of thermal runaway, could have for detonation ignition. Such He cap may partially mix with the underlying carbon-oxygen layers during different phases of the evolution of the white dwarf. Through small-scale one-dimensional spherically-symmetric calculations of detonation ignition in carbon-oxygen matter contaminated with traces of He, we have determined the conditions under which the He presence can allow the formation of a detonation. It turns out that a local contamination by a 10\% mass fraction of He is enough to allow detonation ignition for a peak temperature of just $10^9$~K, provided that the thermal gradient is shallow enough (implying a large enough primer mass). If the He mass fraction is reduced to $X(\mathrm{He})\lesssim0.05$ the peak temperature necessary to detonate rises to $\sim1.8\times10^9$~K. These results might be relevant not only to CDI in PRD models, but also to the more general problem of detonation ignition in any delayed detonation scenario.

Our results show that the PRD scenario of white dwarf delayed detonation is feasible and worth pursuing its study. PRD models may explain many observational features of a wide range of SNIa events, but probably they cannot cover the whole range of energy and radioactive masses deduced from observations. What we have shown in the present work is that the PRD scenario provides a realistic way by which white dwarfs can explode. 
In paper~II we will analyze the explosion phase and compare the final outcome with observational features of typical SNIa.

\acknowledgments

This work has been partially supported by the MEC grants AYA2005-08013-C03, 
AYA2007-66256, by the European Union FEDER funds and by the Generalitat de 
Catalunya, and it is dedicated to the loving memory of Carmen Guil Ruiz

\bibliographystyle{aa}
\bibliography{../../../ebg}

\begin{thebibliography}{71}
\expandafter\ifx\csname natexlab\endcsname\relax\def\natexlab#1{#1}\fi

\bibitem[{{Arnett} \& {Livne}(1994)}]{arn94}
{Arnett}, D. \& {Livne}, E. 1994, \apj, 427, 330

\bibitem[{{Badenes} {et~al.}(2006){Badenes}, {Borkowski}, {Hughes}, {Hwang}, \&
  {Bravo}}]{bad06}
{Badenes}, C., {Borkowski}, K.~J., {Hughes}, J.~P., {Hwang}, U., \& {Bravo}, E.
  2006, \apj, 645, 1373

\bibitem[{{Badenes} {et~al.}(2003){Badenes}, {Bravo}, {Borkowski}, \&
  {Dom{\'{\i}}nguez}}]{bad03}
{Badenes}, C., {Bravo}, E., {Borkowski}, K.~J., \& {Dom{\'{\i}}nguez}, I. 2003,
  \apj, 593, 358

\bibitem[{{Badenes} {et~al.}(2007){Badenes}, {Hughes}, {Bravo}, \&
  {Langer}}]{bad07}
{Badenes}, C., {Hughes}, J.~P., {Bravo}, E., \& {Langer}, N. 2007, \apj, 662,
  472

\bibitem[{{Badenes} {et~al.}(2008){Badenes}, {Hughes}, {Cassam-Chena{\"i}}, \&
  {Bravo}}]{bad08b}
{Badenes}, C., {Hughes}, J.~P., {Cassam-Chena{\"i}}, G., \& {Bravo}, E. 2008,
  \apj, 680, 1149

\bibitem[{{Blinnikov} \& {Khokhlov}(1987)}]{bk87}
{Blinnikov}, S.~I. \& {Khokhlov}, A.~M. 1987, Soviet Astronomy Letters, 13, 364

\bibitem[{{Bravo} {et~al.}(1993){Bravo}, {Dominguez}, {Isern}, {Canal},
  {H\"{o}flich}, \& {Labay}}]{bra93}
{Bravo}, E., {Dominguez}, I., {Isern}, J., {et~al.} 1993, \aap, 269, 187

\bibitem[{{Bravo} \& {Garc{\'{\i}}a-Senz}(2005)}]{bra05b}
{Bravo}, E. \& {Garc{\'{\i}}a-Senz}, D. 2005, in IAU Colloq. 192: Cosmic
  Explosions, On the 10th Anniversary of SN1993J, ed. J.-M. {Marcaide} \& K.~W.
  {Weiler}, 339

\bibitem[{{Bravo} \& {Garc{\'{\i}}a-Senz}(2006)}]{bra06}
{Bravo}, E. \& {Garc{\'{\i}}a-Senz}, D. 2006, \apjl, 642, L157

\bibitem[{{Chamulak} {et~al.}(2008){Chamulak}, {Brown}, {Timmes}, \&
  {Dupczak}}]{ch08}
{Chamulak}, D.~A., {Brown}, E.~F., {Timmes}, F.~X., \& {Dupczak}, K. 2008,
  \apj, 677, 160

\bibitem[{{Colgate} \& {White}(1966)}]{cw66}
{Colgate}, S.~A. \& {White}, R.~H. 1966, \apj, 143, 626

\bibitem[{{Cumming} {et~al.}(1994{\natexlab{a}}){Cumming}, {Meikle}, {Geballe},
  {Royer}, {Hurst}, {Comello}, \& {Dillon}}]{cu94b}
{Cumming}, R.~J., {Meikle}, W.~P.~S., {Geballe}, T.~R., {et~al.}
  1994{\natexlab{a}}, \iaucirc, 5953, 1

\bibitem[{{Cumming} {et~al.}(1994{\natexlab{b}}){Cumming}, {Meikle}, {Geballe},
  {Wall}, {Jenkins}, {Walther}, {Smith}, {Pettini}, {King}, {Martin}, {Shanks},
  {Croom}, {Tanvir}, {Kilmartin}, {Gilmore}, \& {Green}}]{cu94a}
{Cumming}, R.~J., {Meikle}, W.~P.~S., {Geballe}, T.~R., {et~al.}
  1994{\natexlab{b}}, \iaucirc, 5951, 1

\bibitem[{{Dunina-Barkovskaya} \& {Imshennik}(2003)}]{dun03}
{Dunina-Barkovskaya}, N.~V. \& {Imshennik}, V.~S. 2003, Astronomy Letters, 29,
  10

\bibitem[{{Dunina-Barkovskaya} {et~al.}(2001){Dunina-Barkovskaya}, {Imshennik},
  \& {Blinnikov}}]{dun01}
{Dunina-Barkovskaya}, N.~V., {Imshennik}, V.~S., \& {Blinnikov}, S.~I. 2001,
  Astronomy Letters, 27, 353

\bibitem[{{Evrard}(1988)}]{evr88}
{Evrard}, A.~E. 1988, \mnras, 235, 911

\bibitem[{{Fesen} {et~al.}(2007){Fesen}, {H\"oflich}, {Hamilton}, {Hammell},
  {Gerardy}, {Khokhlov}, \& {Wheeler}}]{fes07}
{Fesen}, R., {H\"oflich}, P., {Hamilton}, A., {et~al.} 2007, \apj, 658, 396

\bibitem[{{Gamezo} {et~al.}(2004){Gamezo}, {Khokhlov}, \& {Oran}}]{gam04}
{Gamezo}, V.~N., {Khokhlov}, A.~M., \& {Oran}, E.~S. 2004, Physical Review
  Letters, 92, 211102

\bibitem[{{Gamezo} {et~al.}(2005){Gamezo}, {Khokhlov}, \& {Oran}}]{gam05}
{Gamezo}, V.~N., {Khokhlov}, A.~M., \& {Oran}, E.~S. 2005, \apj, 623, 337

\bibitem[{{Gamezo} {et~al.}(2003){Gamezo}, {Khokhlov}, {Oran}, {Chtchelkanova},
  \& {Rosenberg}}]{gam03}
{Gamezo}, V.~N., {Khokhlov}, A.~M., {Oran}, E.~S., {Chtchelkanova}, A.~Y., \&
  {Rosenberg}, R.~O. 2003, Science, 299, 77

\bibitem[{{Garc{\'{\i}}a-Senz} \& {Bravo}(2005)}]{gar05}
{Garc{\'{\i}}a-Senz}, D. \& {Bravo}, E. 2005, \aap, 430, 585

\bibitem[{{Garcia-Senz} \& {Woosley}(1995)}]{gar95}
{Garcia-Senz}, D. \& {Woosley}, S.~E. 1995, \apj, 454, 895

\bibitem[{{Gerardy} {et~al.}(2007){Gerardy}, {Meikle}, {Kotak}, {H{\"o}flich},
  {Farrah}, {Filippenko}, {Foley}, {Lundqvist}, {Mattila}, {Pozzo},
  {Sollerman}, {Van Dyk}, \& {Wheeler}}]{ger07}
{Gerardy}, C.~L., {Meikle}, W.~P.~S., {Kotak}, R., {et~al.} 2007, \apj, 661,
  995

\bibitem[{{Hachisu} {et~al.}(2008){Hachisu}, {Kato}, \& {Nomoto}}]{hkn08}
{Hachisu}, I., {Kato}, M., \& {Nomoto}, K. 2008, \apj, 679, 1390

\bibitem[{{Han} \& {Podsiadlowski}(2004)}]{han04}
{Han}, Z. \& {Podsiadlowski}, P. 2004, \mnras, 350, 1301

\bibitem[{{Hatano} {et~al.}(1999){Hatano}, {Branch}, {Fisher}, {Baron}, \&
  {Filippenko}}]{ha99}
{Hatano}, K., {Branch}, D., {Fisher}, A., {Baron}, E., \& {Filippenko}, A.~V.
  1999, \apj, 525, 881

\bibitem[{{Hillebrandt} \& {Niemeyer}(2000)}]{hil00}
{Hillebrandt}, W. \& {Niemeyer}, J.~C. 2000, \araa, 38, 191

\bibitem[{{H{\"o}flich} {et~al.}(2003){H{\"o}flich}, {Gerardy}, {Linder}, \&
  {et al.}}]{hoe03}
{H{\"o}flich}, P., {Gerardy}, C., {Linder}, E., \& {et al.} 2003, in Lecture
  Notes in Physics, Berlin Springer Verlag, Vol. 635, Stellar Candles for the
  Extragalactic Distance Scale, ed. D.~{Alloin} \& W.~{Gieren}, 203--227

\bibitem[{{H\"{o}flich} \& {Khokhlov}(1996)}]{hk96}
{H\"{o}flich}, P. \& {Khokhlov}, A. 1996, \apj, 457, 500

\bibitem[{{H\"{o}flich} {et~al.}(1995){H\"{o}flich}, {Khokhlov}, \&
  {Wheeler}}]{hoe95}
{H\"{o}flich}, P., {Khokhlov}, A.~M., \& {Wheeler}, J.~C. 1995, \apj, 444, 831

\bibitem[{{Howell} {et~al.}(2001){Howell}, {H{\"o}flich}, {Wang}, \&
  {Wheeler}}]{how01}
{Howell}, D.~A., {H{\"o}flich}, P., {Wang}, L., \& {Wheeler}, J.~C. 2001, \apj,
  556, 302

\bibitem[{{Ivanova} {et~al.}(1974){Ivanova}, {Imshennik}, \&
  {Chechetkin}}]{iva74}
{Ivanova}, L.~N., {Imshennik}, V.~S., \& {Chechetkin}, V.~M. 1974, \apss, 31,
  497

\bibitem[{{Jordan} {et~al.}(2008){Jordan}, {Fisher}, {Townsley}, {Calder},
  {Graziani}, {Asida}, {Lamb}, \& {Truran}}]{jor07}
{Jordan}, G.~I., {Fisher}, R., {Townsley}, D., {et~al.} 2008, \apj, 681, 1448

\bibitem[{{Khokhlov} {et~al.}(1993){Khokhlov}, {Mueller}, \&
  {H\"{o}flich}}]{kho93}
{Khokhlov}, A., {Mueller}, E., \& {H\"{o}flich}, P. 1993, \aap, 270, 223

\bibitem[{{Khokhlov}(1989)}]{kho89}
{Khokhlov}, A.~M. 1989, \mnras, 239, 785

\bibitem[{{Khokhlov}(1991)}]{kho91}
{Khokhlov}, A.~M. 1991, \aap, 245, 114

\bibitem[{{Khokhlov} {et~al.}(1997){Khokhlov}, {Oran}, \& {Wheeler}}]{kho97}
{Khokhlov}, A.~M., {Oran}, E.~S., \& {Wheeler}, J.~C. 1997, \apj, 478, 678

\bibitem[{{Kitsionas} \& {Whitworth}(2002)}]{kit02}
{Kitsionas}, S. \& {Whitworth}, A.~P. 2002, \mnras, 330, 129

\bibitem[{{Kuhlen} {et~al.}(2006){Kuhlen}, {Woosley}, \& {Glatzmaier}}]{kuh06}
{Kuhlen}, M., {Woosley}, S.~E., \& {Glatzmaier}, G.~A. 2006, \apj, 640, 407

\bibitem[{{Langer} {et~al.}(2000){Langer}, {Deutschmann}, {Wellstein}, \&
  {H{\"o}flich}}]{lan00}
{Langer}, N., {Deutschmann}, A., {Wellstein}, S., \& {H{\"o}flich}, P. 2000,
  \aap, 362, 1046

\bibitem[{{Lisewski} {et~al.}(2000){Lisewski}, {Hillebrandt}, \&
  {Woosley}}]{lis00}
{Lisewski}, A.~M., {Hillebrandt}, W., \& {Woosley}, S.~E. 2000, \apj, 538, 831

\bibitem[{{Livne} {et~al.}(2005){Livne}, {Asida}, \& {H{\"o}flich}}]{liv05}
{Livne}, E., {Asida}, S.~M., \& {H{\"o}flich}, P. 2005, \apj, 632, 443

\bibitem[{{Mazzali} \& {Lucy}(1998)}]{ml98}
{Mazzali}, P.~A. \& {Lucy}, L.~B. 1998, \mnras, 295, 428

\bibitem[{{Niemeyer}(1999)}]{nie99}
{Niemeyer}, J.~C. 1999, \apjl, 523, L57

\bibitem[{{Niemeyer} \& {Woosley}(1997)}]{nie97}
{Niemeyer}, J.~C. \& {Woosley}, S.~E. 1997, \apj, 475, 740

\bibitem[{{Nomoto} {et~al.}(2003){Nomoto}, {Uenishi}, {Kobayashi}, {Umeda},
  {Ohkubo}, {Hachisu}, \& {Kato}}]{no03}
{Nomoto}, K., {Uenishi}, T., {Kobayashi}, C., {et~al.} 2003, in From Twilight
  to Highlight: The Physics of Supernovae, ed. W.~{Hillebrandt} \&
  B.~{Leibundgut}, 115

\bibitem[{{Pan} {et~al.}(2008){Pan}, {Wheeler}, \& {Scalo}}]{pan08}
{Pan}, L., {Wheeler}, J.~C., \& {Scalo}, J. 2008, \apj, 681, 470

\bibitem[{{Perlmutter} {et~al.}(1998){Perlmutter}, {Aldering}, {della Valle},
  {Deustua}, {Ellis}, {Fabbro}, {Fruchter}, {Goldhaber}, {Groom}, {Hook},
  {Kim}, {Kim}, {Knop}, {Lidman}, {McMahon}, {Nugent}, {Pain}, {Panagia},
  {Pennypacker}, {Ruiz-Lapuente}, {Schaefer}, \& {Walton}}]{per98}
{Perlmutter}, S., {Aldering}, G., {della Valle}, M., {et~al.} 1998, \nat, 391,
  51

\bibitem[{{Piersanti} {et~al.}(2003){Piersanti}, {Gagliardi}, {Iben}, \&
  {Tornamb{\'e}}}]{pie03}
{Piersanti}, L., {Gagliardi}, S., {Iben}, I.~J., \& {Tornamb{\'e}}, A. 2003,
  \apj, 598, 1229

\bibitem[{{Pignata} {et~al.}(2008){Pignata}, {Benetti}, {Mazzali}, {Kotak},
  {Patat}, {Meikle}, {Stehle}, {Leibundgut}, {Suntzeff}, {Buson}, {Cappellaro},
  {Clocchiatti}, {Hamuy}, {Maza}, {Mendez}, {Ruiz-Lapuente}, {Salvo},
  {Schmidt}, {Turatto}, \& {Hillebrandt}}]{pi08}
{Pignata}, G., {Benetti}, S., {Mazzali}, P.~A., {et~al.} 2008, \mnras, 388, 971

\bibitem[{{Piro} \& {Bildsten}(2008)}]{pb08}
{Piro}, A.~L. \& {Bildsten}, L. 2008, \apj, 673, 1009

\bibitem[{{Plewa}(2007)}]{pl07}
{Plewa}, T. 2007, \apj, 657, 942

\bibitem[{{Plewa} {et~al.}(2004){Plewa}, {Calder}, \& {Lamb}}]{pl04}
{Plewa}, T., {Calder}, A.~C., \& {Lamb}, D.~Q. 2004, \apjl, 612, L37

\bibitem[{{Quimby} {et~al.}(2007){Quimby}, {H{\"o}flich}, \& {Wheeler}}]{qui07}
{Quimby}, R., {H{\"o}flich}, P., \& {Wheeler}, J.~C. 2007, \apj, 666, 1083

\bibitem[{{Reinecke} {et~al.}(2002){Reinecke}, {Hillebrandt}, \&
  {Niemeyer}}]{rei02b}
{Reinecke}, M., {Hillebrandt}, W., \& {Niemeyer}, J.~C. 2002, \aap, 391, 1167

\bibitem[{{Rest} {et~al.}(2008){Rest}, {Matheson}, {Blondin}, {Bergmann},
  {Welch}, {Suntzeff}, {Smith}, {Olsen}, {Prieto}, {Garg}, {Challis}, {Stubbs},
  {Hicken}, {Modjaz}, {Wood-Vasey}, {Zenteno}, {Damke}, {Newman}, {Huber},
  {Cook}, {Nikolaev}, {Becker}, {Miceli}, {Covarrubias}, {Morelli}, {Pignata},
  {Clocchiatti}, {Minniti}, \& {Foley}}]{res08}
{Rest}, A., {Matheson}, T., {Blondin}, S., {et~al.} 2008, \apj, 680, 1137

\bibitem[{{Riess} {et~al.}(2001){Riess}, {Nugent}, {Gilliland}, {Schmidt},
  {Tonry}, {Dickinson}, {Thompson}, {Budav{\'a}ri}, {Casertano}, {Evans},
  {Filippenko}, {Livio}, {Sanders}, {Shapley}, {Spinrad}, {Steidel}, {Stern},
  {Surace}, \& {Veilleux}}]{rie01}
{Riess}, A.~G., {Nugent}, P.~E., {Gilliland}, R.~L., {et~al.} 2001, \apj, 560,
  49

\bibitem[{{R{\"o}pke} {et~al.}(2004){R{\"o}pke}, {Hillebrandt}, \&
  {Niemeyer}}]{rop04}
{R{\"o}pke}, F.~K., {Hillebrandt}, W., \& {Niemeyer}, J.~C. 2004, \aap, 421,
  783

\bibitem[{{R{\"o}pke} {et~al.}(2007{\natexlab{a}}){R{\"o}pke}, {Hillebrandt},
  {Schmidt}, {Niemeyer}, {Blinnikov}, \& {Mazzali}}]{rop07c}
{R{\"o}pke}, F.~K., {Hillebrandt}, W., {Schmidt}, W., {et~al.}
  2007{\natexlab{a}}, \apj, 668, 1132

\bibitem[{{R{\"o}pke} \& {Niemeyer}(2007)}]{rop07}
{R{\"o}pke}, F.~K. \& {Niemeyer}, J.~C. 2007, \aap, 464, 683

\bibitem[{{R{\"o}pke} {et~al.}(2007{\natexlab{b}}){R{\"o}pke}, {Woosley}, \&
  {Hillebrandt}}]{rwh07}
{R{\"o}pke}, F.~K., {Woosley}, S.~E., \& {Hillebrandt}, W. 2007{\natexlab{b}},
  \apj, 660, 1344

\bibitem[{{Schmidt} {et~al.}(1998){Schmidt}, {Suntzeff}, {Phillips},
  {Schommer}, {Clocchiatti}, {Kirshner}, {Garnavich}, {Challis}, {Leibundgut},
  {Spyromilio}, {Riess}, {Filippenko}, {Hamuy}, {Smith}, {Hogan}, {Stubbs},
  {Diercks}, {Reiss}, {Gilliland}, {Tonry}, {Maza}, {Dressler}, {Walsh}, \&
  {Ciardullo}}]{sch98}
{Schmidt}, B.~P., {Suntzeff}, N.~B., {Phillips}, M.~M., {et~al.} 1998, \apj,
  507, 46

\bibitem[{{Townsley} {et~al.}(2007){Townsley}, {Calder}, {Asida}, {Seitenzahl},
  {Peng}, {Vladimirova}, {Lamb}, \& {Truran}}]{tow07}
{Townsley}, D.~M., {Calder}, A.~C., {Asida}, S.~M., {et~al.} 2007, \apj, 668,
  1118

\bibitem[{{Wheeler} {et~al.}(1998){Wheeler}, {H\"{o}flich}, {Harkness}, \&
  {Spyromilio}}]{wh98}
{Wheeler}, J.~C., {H\"{o}flich}, P., {Harkness}, R.~P., \& {Spyromilio}, J.
  1998, \apj, 496, 908

\bibitem[{{Woosley}(1990)}]{woo90}
{Woosley}, S.~E. 1990, in Supernovae, ed. A.~G. {Petschek}, 182--212

\bibitem[{{Woosley}(2007)}]{woo07b}
{Woosley}, S.~E. 2007, \apj, 668, 1109

\bibitem[{{Woosley} \& {Weaver}(1994)}]{woo94}
{Woosley}, S.~E. \& {Weaver}, T.~A. 1994, in Supernovae, ed. S.~A. {Bludman},
  R.~{Mochkovitch}, \& J.~{Zinn-Justin}, 63

\bibitem[{{Woosley} {et~al.}(2004){Woosley}, {Wunsch}, \& {Kuhlen}}]{woo04}
{Woosley}, S.~E., {Wunsch}, S., \& {Kuhlen}, M. 2004, \apj, 607, 921

\bibitem[{{Wunsch} \& {Woosley}(2004)}]{wun04}
{Wunsch}, S. \& {Woosley}, S.~E. 2004, \apj, 616, 1102

\bibitem[{{Zingale} \& {Dursi}(2007)}]{zin07}
{Zingale}, M. \& {Dursi}, L.~J. 2007, \apj, 656, 333

\bibitem[{{Zingale} {et~al.}(2005){Zingale}, {Woosley}, {Bell}, {Day}, \&
  {Rendleman}}]{zin05}
{Zingale}, M., {Woosley}, S.~E., {Bell}, J.~B., {Day}, M.~S., \& {Rendleman},
  C.~A. 2005, Journal of Physics Conference Series, 16, 405

\end{thebibliography}

\clearpage
\centering
\begin{deluxetable}{rrrr}
\tabletypesize{\scriptsize}
\tablecaption{Initial size and position of the hot bubbles\tablenotemark{a}}
\tablecolumns{4}
\tablewidth{0pt}
\tablehead{
\colhead{$x_\mathrm{c}$} &
\colhead{$y_\mathrm{c}$} &
\colhead{$z_\mathrm{c}$} &
\colhead{$M_\mathrm{b}$} \\
(km) & (km) & (km) & (g) \\
}
\startdata
 -56.5 &   74.5 &   87.0 & $1.59\times10^{30}$ \\
 101.3 &  -87.7 &  -52.2 & $1.48\times10^{30}$ \\
 106.5 &   41.3 &   47.2 & $1.53\times10^{30}$ \\
 115.9 &  180.3 & -195.0 & $1.09\times10^{30}$ \\
-145.8 &  -83.4 &   35.8 & $1.44\times10^{30}$ \\
 275.9 & -101.5 &  -80.4 & $0.87\times10^{30}$ \\
  32.4 &  239.6 &  -61.6 & $1.35\times10^{30}$ \\
\enddata
\tablenotetext{a}{In this table $x_\mathrm{c}$, $y_\mathrm{c}$  and $z_\mathrm{c}$ are the coordinates of the center of the bubbles, while $M_\mathrm{b}$ gives the mass of each bubble}
\label{tab0}
\end{deluxetable}

\clearpage
\centering
\begin{deluxetable}{lccc}
\tabletypesize{\scriptsize}
\tablecaption{Physical conditions of shocked matter}
\tablecolumns{4}
\tablewidth{0pt}
\tablehead{
\colhead{Model} &
\colhead{DF11} &
\colhead{DF18} &
\colhead{DF29} \\
}
\startdata
$M_\mathrm{defl}$ (M$_\odot$) & 0.11 & 0.18 & 0.29 \\
$t_\mathrm{acc}$ (s)\tablenotemark{a} & 4.00 & 7.18 & 12.17 \\
$\gamma M^2$ & 12 & 12 & 9 \\
$\rho_\mathrm{shock}$ (g cm$^{-3}$) & $5.2\times10^5$ & $3.3\times10^5$ & $1.4\times10^5$ \\
$T_\mathrm{shock}$ (K) & $8.1\times10^8$ & $10^9$ & $4.5\times10^8$ \\
$\varepsilon_\mathrm{mech}$ (erg~s$^{-1}$) & $1.0\times10^{49}$ & $10^{49}$ & $5\times10^{48}$ \\
$M_\mathrm{pri}$ (g) & $5.2\times10^{30}$ & $5.2\times10^{30}$ & $2.6\times10^{30}$ \\
$\tau_\mathrm{hydro}$ (s)\tablenotemark{b} & 0.62 & 0.78 & 1.21 \\
$M_\mathrm{deto}$ (M$_\odot$)\tablenotemark{c} & 1.29 & 1.10 & 0.38 \\
$t_\mathrm{deto}$ (s)\tablenotemark{d} & 4.3 & 7.5 & - \\
\enddata
\tablenotetext{a}{Time at which the accretion shock is formed}
\tablenotetext{b}{Hydrodynamic timescale just behind the accretion shock}
\tablenotetext{c}{ {Lagrangian} mass at which detonation ignition conditions would be achieved. See Sect.~\ref{phys}}
\tablenotetext{d}{Time at which detonation ignition conditions are achieved}
\label{tab1}
\end{deluxetable}

\clearpage
\centering
\begin{deluxetable}{ccc}
\tabletypesize{\scriptsize}
\tablecaption{Summary of detonation initiation conditions in C-O matter\tablenotemark{a}}
\tablecolumns{3}
\tablewidth{0pt}
\tablehead{
\colhead{$\rho$} &
\colhead{$T_\mathrm{c}$} &
\colhead{$M_\mathrm{pri}$} \\
(g~cm$^{-3}$) & ($10^9$~K) & (g) \\
}
\startdata
$3\times10^6$ & 2.3 & $2\times10^{28}$ \\
$10^7$ & 1.9 & $1.5\times10^{27}$ \\
$10^7$ & 2 & $1.1\times10^{24}$ \\
$10^7$ & 2.2 & $2\times10^{25}$ \\
$10^7$ & 2.8 & $2.5\times10^{23}$ \\
$10^7$ & 3.5 & $2\times10^{23}$ \\
$2\times10^7$ & 2 & $2.9\times10^{21}$ \\
$3\times10^7$ & 2 & $4.9\times10^{20}$ \\
$3\times10^7$ & 5.2 & $2\times10^{19}$ \\
$5\times10^7$ & 2 & $1.1\times10^{19}$ \\
$10^8$ & 2 & $5.5\times10^{17}$ \\
$10^8$ & 6.2 & $2\times10^{15}$ \\
\enddata
\tablenotetext{a}{Compiled from \citet{arn94,nie97,rwh07}}
\label{tab2}
\end{deluxetable}

\clearpage
\centering
\begin{deluxetable}{lcccc}
\tabletypesize{\scriptsize}
\tablecaption{Maximum temperature reached with different mass resolutions}
\tablecolumns{5}
\tablewidth{0pt}
\tablehead{
\colhead{Model} &
\colhead{$\rho_\mathrm{max}$} &
\colhead{$T_\mathrm{max}$} &
\colhead{$\Delta m_\mathrm{min}$\tablenotemark{a}} &
\colhead{$\Delta x_\mathrm{min}$\tablenotemark{b}} \\
 & (g~cm$^{-3}$) & ($10^9$~K) & (g) & (km) \\
}
\startdata
DF18      & $4.6\times10^6$ & 2.40 & $1.1\times10^{28}$ & 83  \\
DF18sp2   & $4.5\times10^6$ & 2.36 & $5.5\times10^{27}$ & 66  \\
DF18sp4   & $2.7\times10^6$ & 2.76 & $2.8\times10^{27}$ & 63  \\
DF18sp10  & $4.6\times10^6$ & 2.74 & $1.1\times10^{27}$ & 39  \\
DF18sp100 & $3.7\times10^6$ & 3.70 & $1.1\times10^{26}$ & 19  \\
DF29      & $1.3\times10^6$ & 1.42 & $1.1\times10^{28}$ & 126 \\
DF29sp10  & $2.6\times10^6$ & 1.91 & $1.1\times10^{27}$ & 47  \\
DF29sp100 & $1.7\times10^6$ & 1.95 & $1.1\times10^{26}$ & 25  \\
\enddata
\tablenotetext{a}{Maximum mass resolution}
\tablenotetext{b}{Maximum spatial resolution, given by $\Delta x_\mathrm{min} = \left(3\Delta m_\mathrm{min}/4\pi\rho_\mathrm{max}\right)^{1/3}$}
\label{tab3}
\end{deluxetable}

\clearpage
\centering
\begin{deluxetable}{lccccr}
\tabletypesize{\scriptsize}
\tablecaption{Detonation initiation in C-O matter with traces of He}
\tablecolumns{6}
\tablewidth{0pt}
\tablehead{
\colhead{$T$} & \colhead{$M_\mathrm{pri}$\tablenotemark{a}} & \colhead{$X(\mathrm{He})$\tablenotemark{b}} & \colhead{He-distribution\tablenotemark{c}} & \colhead{Detonation?} & $\nabla T$ \\
($10^9$~K) & (g) & & & & (K~cm$^{-1}$)\\
}
\startdata
1.0 & $3\times10^{29}$ & 0.10 & primer & y & 28 \\
1.4 & $3\times10^{25}$ & 0.10 & unif. & y & 830 \\
1.4 & $3\times10^{27}$ & 0.02 & primer & n & 180 \\
1.4 & $3\times10^{27}$ & 0.05 & primer & n & 180 \\
1.4 & $3\times10^{27}$ & 0.10 & primer & n & 180 \\
1.4 & $3\times10^{27}$ & 0.20 & primer & n & 180 \\
1.4 & $3\times10^{27}$ & 0.10 & unif. & y & 180 \\
1.4 & $3\times10^{29}$ & 0.05 & primer & n & 39 \\
1.4 & $3\times10^{29}$ & 0.10 & primer & y & 39 \\
1.8 & $3\times10^{27}$ & 0.05 & unif. & n & 230 \\
1.8 & $3\times10^{29}$ & 0.05 & primer & y & 50 \\
\enddata
\tablenotetext{a}{In all calculations the density was $\rho=1.5\times10^6$~g~cm$^{-3}$}
\tablenotetext{b}{He mass fraction}
\tablenotetext{c}{Primer: He concentrated in the primer. Unif.: He uniformly distributed through the whole microsphere}
\label{tab4}
\end{deluxetable}

\clearpage

\end{document}